\documentclass[aip,jap,amsmath, amssymb,reprint,floatfix]{revtex4-2}
\usepackage{graphicx}

\usepackage{dcolumn}
\usepackage{bm}
\usepackage[utf8]{inputenc}

\usepackage{mathptmx}
\usepackage{etoolbox}
\usepackage{xcolor}

\usepackage{ulem}

\makeatletter
\def\@email#1#2{%
 \endgroup
 \patchcmd{\titleblock@produce}
  {\frontmatter@RRAPformat}
  {\frontmatter@RRAPformat{\produce@RRAP{*#1\href{mailto:#2}{#2}}}\frontmatter@RRAPformat}
  {}{}
}%
\makeatother

\begin{document}

\title{Energetics and thermodynamics of bilayer rectangular artificial spin ices}

\author{Gabriel A. Oliveira}
\affiliation{Departamento de F\'{i}sica, Universidade Federal de Vi\c{c}osa, Vi\c{c}osa,	36570-900, Minas Gerais, Brazil}
\email{gabriel.oliveira7@ufv.br}

\author{Winder A. Moura-Melo}
\affiliation{Departamento de F\'{i}sica, Universidade Federal de Vi\c{c}osa, Vi\c{c}osa,	36570-900, Minas Gerais, Brazil}

\author{Afranio R. Pereira}
\affiliation{Departamento de F\'{i}sica, Universidade Federal de Vi\c{c}osa, Vi\c{c}osa, 36570-900, Minas Gerais, Brazil}

\author{Fabio S. Nascimento}
\affiliation{Centro de Forma\c{c}ao de Professores, Universidade Federal do Rec\^{o}ncavo da Bahia, Amargosa, 45300-000, Bahia, Brazil}
\date{\today}

\begin{abstract}
{Bilayer rectangular artificial spin ices (BRASIs) with distinct aspect ratios, $\gamma$, are considered. Namely, we investigate how the underlying geometry modifies the interaction between two rectangular artificial spin ice layers separated by a height offset, $h$, whenever compared to the square case. Actually, rectangular layers interact by means of a Buckingham-like potential, whereas in the square case, one has an algebraic (van der Walls-like) interaction. In addition, Moiré patterns for BRASIs are less definite than for the square bilayer. We also deal with their basic thermodynamics, showing the behavior of the specific heat as a function of temperature and $\gamma$ for a number of height offsets.}
\end{abstract}

\maketitle

\section{Introduction and Motivation}
Magnetic frustrated systems\cite{MoessnerToday} have received a great deal of effort in recent years. Besides their importance as frameworks for emerging fundamental phenomena, there is also a number of promising applications ranging from magnetic sensors to spintronics. Among such systems, those termed artificial spin ices (ASIs) \cite{Wang-Schiffer-ASI-Nature-2006,Review1,Shakti,Nisoli-17,Review2,Review3} have deserved considerable attention, namely because geometrical frustration may be controlled on demand by choosing specific arrangements of the magnetic dipoles composing the system \cite{Design}. Such dipoles come from magnets with strong Ising shape anisotropy \cite{Wysin1,Wysin2} (e.g., elongated magnetic bars or islands). These islands may be very tiny, having  a few to several nanometers (nanoislands)\cite{Wang-Schiffer-ASI-Nature-2006} or they may be macroscopic as well, with several millimeters or even centimeters\cite{Paula-Mellado-PRL-2012,Leon-JMMM-2013,Macro-ASI-JMMM-2020,Hamilton24}. Although energy scales are quite distinct, the main physics taking place at nanosized ASIs is essentially the same observed at the macroscopic scale \cite{Macro-ASI-JMMM-2020}, whenever the underlying geometry remains the same.\\

However, keeping the size dimensions but changing the geometry, two different arrangements will exhibit distinct phenomena as consequences of their intrinsic geometrical frustration. For instance, both square and rectangular ASIs have ground states respecting ice rule at every vertex: 2 dipoles point in, whereas the other 2 point out (see Fig. 1). However, their excitation spectra above the ground state present differences: in the square geometry, one has monopole pairs joined by energetic strings, which may be faced as condensed matter analogues of Nambu monopoles \cite{Nambu74,Volovik,Silva-2013,Cumings-2014,Marrows-2019}. Thus, the dissociation of a monopole pair and the subsequent spatial separation of their opposite poles have an energetic cost\cite{JAP-2009,PRB-2010,Morgan-2011}. If the square arrangement is suitably stretched along one side to produce a rectangular ASI (with lattice parameters $a$ and $b$), then string tension diminishes, eventually vanishing at a specific aspect ratio, $\gamma_c=a/b=\sqrt{3}$, allowing the emergence of monopoles as essentially free excitations \cite{Rectangular-2012,RectangularASI-2,JAP2024,APL-2024}.\\

Recently, stacking of two layers of square ASIs separated by a height offset, $h$, has been investigated, and the system's energetic and excitations have been discussed in Ref. \cite{Bilayer-Fabio-21}. Whenever the bilayer compound is in its true ground state, then the layers attract each other with a power-law force resembling a van der Waals-like interaction. The presence of excitations (Nambu monopoles) in one or both layers renders a switching between attraction and repulsion, depending on $h$ and charges separation \cite{Bilayer-Fabio-21}. Additionally, whenever the layers are  twisted each other, unusual magnetic ordering appears for special angles defining lateral spin super-lattices in both square and pinwheel arrangements \cite{Bob-22}. In turn, stacked magnetic systems, such as ASIs, moir\'{e} patterns of dipoles \cite{Mellado-2024}, and twisted van der Waals magnetic layers \cite{Balents-2020} may bring relevant ideas for the emerging field of twistronics. At some extent, the stacking of two or more layers is one of the most feasible routes to achieve three-dimensional ASI systems \cite{Berchialla-APL-125-2024}. In addition, such stacked systems may bring novelties on the monopole magnetohydrodynamics and magnetosonic wave instability predicted to occur in single layer ASI \cite{Banerjee-PRB-2025}.

Here, we investigate bilayer rectangular artificial spin ices (BRASIs). Firstly, let us recall that in a single layer of rectangular ASI, three different ground states show up, depending on the aspect ratio, $\gamma\equiv a/b$. At the critical value, $\gamma_{c} = \sqrt3$, $t_1$- or $t_2$-type vertices (see Fig. 1) are energetically equivalent and one has a degenerate ground state. For $\gamma<\gamma_c$, only $t_1$-type vertices appear, whereas only $t_2$-type takes place if $\gamma>\gamma_c$. Here, we shall systematically study the three representative cases of rectangular arrangement, $\gamma= \sqrt2, \sqrt3, \sqrt4$, which capture all the main physical properties and nuances of a single- or a bilayer rectangular ASI system. Actually, while square layers attract (or repel) each other by means of an algebraic van der Waals-like force\cite{Bilayer-Fabio-21}, rectangular layers coupling is augmented by an exponential term, making the interaction resemble a Buckingham-type potential\cite{Buckingham}. Such a potential was originally proposed to study the equation of state for gaseous helium, neon and argon, in such a way that the repulsion (exponential-like) term is due to the interpenetration of the closed electron shells \cite{Buckingham}. In general grounds, this suggests the stacking of ASI layers as a magnetic analogue of molecular interaction, offering a platform to study some aspects of such an interaction in a controllable way, namely those concerning dipole-dipole coupling between molecules. 

Finally, we also study the thermodynamics of this system by considering the behavior of the specific heat, discussing the main differences between BRASI and the square bilayer case.

\section{Model and Methods}

Dipole-dipole interaction is the starting point to investigate the energetics and excitation spectrum coming about from every artificial spin ice system, as follows:

\begin{equation}
    H = D \sum_{i>j}\left[\frac{\hat{e}_i\cdot\hat{e}_j} {r_{ij}^3} - 3\frac{(\hat{e}_i\cdot \vec{r_{ij}})(\hat{e}_j\cdot \vec{r_{ij}})}{r_{ij}^5}\right]s_i s_j,
\end{equation}
where $D=\mu_0 \mu^2/4\pi$ is the dipole-dipole strength constant, while the sum is carried out over all magnetic dipoles lying on sites $(i,j)$, spatially separated by $r_{ij}$, throughout the whole system comprising the both layers with a height offset $h$ between them, see Fig. 1. In addition, $\hat{e}_i$ is the unity vector of the local Ising lattice axis, $r_{ij}$ is the distance between dipole $i$ and dipole $j$ while $s_i = \pm 1$ accounts for the direction of the dipole moment at each site, since due to shape anisotropy, elongated nanoislands behave as effective Ising-like dipoles. Each rectangular ASI layer has two lattice spacing parameters, $a$ and $b$, along $x$ and $y$ directions, so that the aspect ratio, $\gamma\equiv a/b$, measures how much a layer is stretched along a given direction($\gamma=1$ recovers the square case). 
\begin{figure}  
\includegraphics[width=0.99\linewidth]{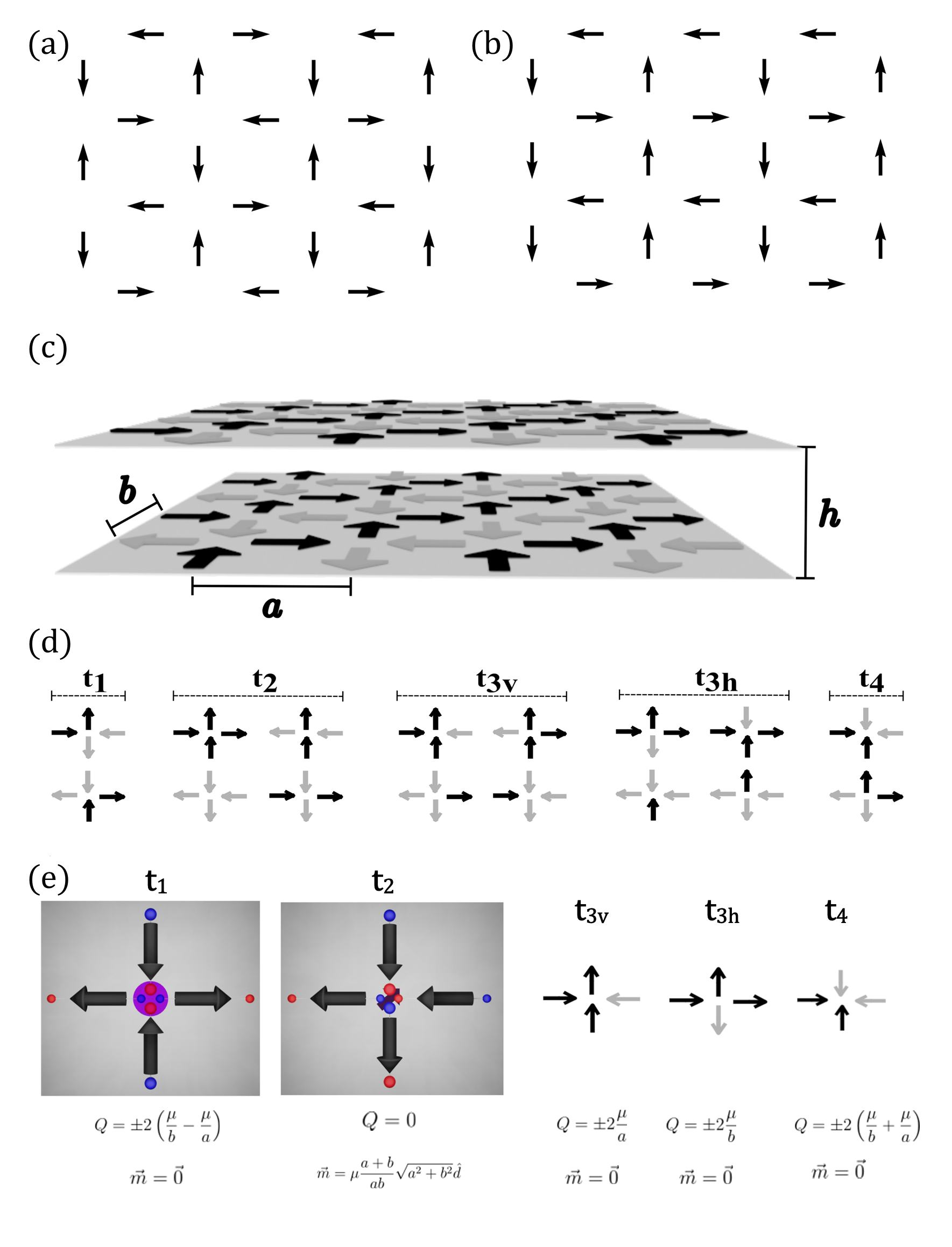}
\caption{\label{fig:fig1} Panels {\bf a)} and {\bf b)} show typical $S1$ and $S2$ states, composed by $t_1$- and $t_2$-vertex, respectively. Namely, note that  $t_1$-vertex appear alternating residual charges, so that the whole system remains magnetically neutral; {\bf c)} BRASI system composed by two identical layers disposed parallel each other with a height offset, $h$, between them. The layers may be twisted each other by an angle $\varphi$; {\bf d)} The five vertex types for all possible 16 arrangements of 4 dipoles per vertex; {\bf e)} Residual charge, $Q$, or net diagonal magnetic moment, $\vec{m}$, locally appear at a vertex as the layer is stretched along one direction (for further details, see Appendix A and Refs.\cite{Rectangular-2012,RectangularASI-2}). }
\end{figure}
Let us briefly recall what occurs in the square case\cite{Bilayer-Fabio-21}, $\gamma=1$: the ground state is achieved when all the vertices of both layers exhibit the $t_1$ configuration, which obeys the ice rule and carries vanishing residual charge and magnetic moment. In addition, the layers experience an attraction between them according to a van der Waals-like force. Such a force may become repulsive whenever suitable monopole excitation comes about in one or both layers.

Departing from the square to a rectangular geometry, a number of novelties appear even for a single layer. For instance: i) $t_1$, $t_3$, and $t_4$ vertices now carry residual magnetic charge, whose magnitude increases with lattice stretching, whereas $t_2$ bears residual magnetic moment, see Fig. 1; ii) At $\gamma=a/b=\sqrt3\equiv \gamma_c$ $t_1$ and $t_2$ vertices have the same energy, yielding to degenerate ground state and vanishing string tension. At this critical aspect ratio, $\gamma_c$, monopole excitations become practically free to move throughout the rectangular ASI\cite{Rectangular-2012,RectangularASI-2,JAP2024,APL-2024}; iii) Below (above) $\gamma_c$, ground state is composed solely by $t_1$ ($t_2$) vertices, termed S1 (S2), as depicted in Fig. 1. Note also that these ground states are 2- and 8-fold degenerate, respectively. Although a spin ice may carry a net magnetic moment (residual  magnetization along the diagonal, $\vec{m}$), magnetic charges cancel each other, so that the whole system remains magnetically neutral. This holds even in the presence of magnetic monopole excitation, carried out by $t_3$ and $t_4$ vertices, since they show up in pairs bound by strings.

Consider now two identical rectangular ASI layers disposed with a fixed height offset, $h$, between them, as depicted in Fig. 1c. In our simulation, each layer comprises 841 vertices and 1740 nanoislands/dipoles. Let $b\equiv 1$ fixed, whereas the aspect ratio $\gamma=a/b \in [1,2]$, ranging over a number of rectangular arrangements, including the critical case, $\gamma_c=\sqrt3$. Ground states for each system are obtained by Monte Carlo steps firstly applied to each layer and then used as an input for the bilayer system. As we shall see, the coupling energy between the layers depends on $h$ and their respective configuration, as well. A number of these results are presented in Fig. 2, along with Table 1, in the next section.
\section{Energetics of BRASI systems}
At first glance, one expects that as $\gamma$ increases, the interactions between layers should diminish once dipole spatial separation becomes larger. However, our simulation shows that energetics tends to become stronger as $\gamma$ is enlarged. Actually, the net interaction in every BRASI system comprises not only the coupling between dipoles; rather, it also takes into account the contribution due to residual charges, whose magnitude increases with $\gamma$ (in general, even much more than residual magnetic momenta. See Appendix for further details). In addition to the relative strength, BRASIs bring about other novelties not revealed by  bilayer square ASI. Indeed, the layers in every BRASI system interact by means of a Buckingham-like potential, as below:
\begin{equation}\label{Potencial}
    {v}(h) = Ah^B + Ce^{Dh}.
\end{equation}
Note that this is the reduced potential (potential strength per vertex). So, its derivative gives the reduced force (force/vertex). Then, for the whole system composed of $N$ vertices, one readily obtains the net potential as $V(h)=N v(h)$.

\begin{figure}[t]
\includegraphics[width= 0.99\linewidth]{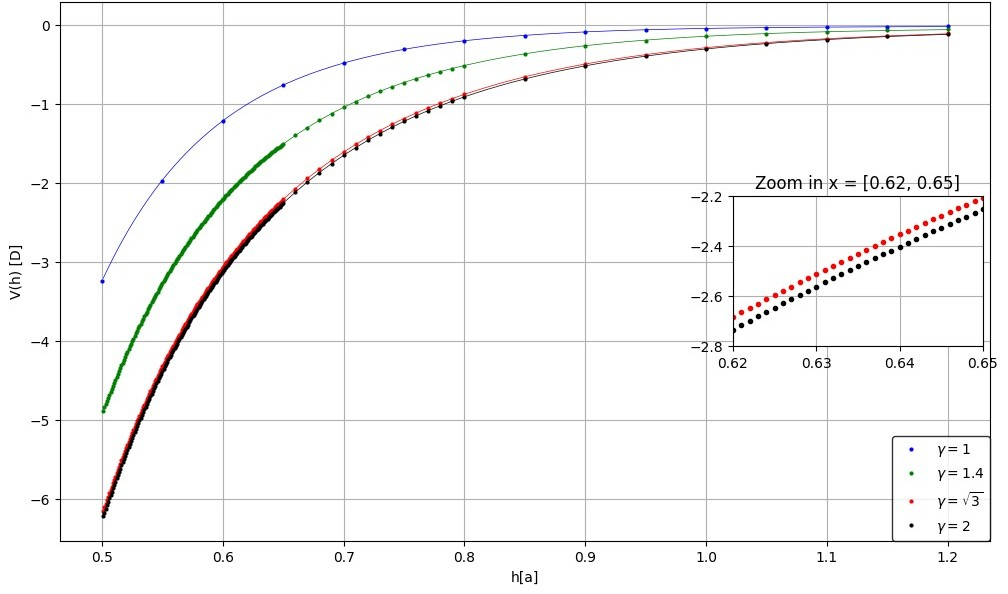} 
\caption{How energy density (energy per vertex) behaves with the height offset, $h$. The layers interact with each other according to a Buckingham-like potential, Eq.(2). Attraction (or repulsion) between them depends on the respective configuration of the bilayer system. Above, we have depicted only the attractive cases, $\nu(h)<0$, by choosing $A<0$  in Table 1, which means that the layers have opposite dipole configurations.} 
\end{figure}

As depicted in Fig. 2 (see also Table 1), the interaction between the layers is quite sensitive to the $\gamma$-parameter below $\gamma_c$, while it remains practically unchanged for $\gamma_c\leq \gamma\leq 2$ (see Appendix). As in the case of the square geometry, we can have an attractive or repulsive interaction if the layers have identical dipole configuration or an exact opposite dipole configuration (coined GS1-GS1 or GS1-GS2 in previous works), respectively.

Indeed, in square geometry, the layers interact with each other through an algebraic van der Waals-type potential ($C=D=0$). However, a slight stretch above $\gamma=1$ changes the scenario, and an extra attractive exponential-like term appears. Thus, for every rectangular ASI, $\gamma>1$, layers interaction must be augmented by a Buckingham-like potential, as expressed in eq. (\ref{Potencial}). Historically, Buckingham proposed it as a simplification of the Lennard-Jones potential in a theoretical study of the equation of state for noble gaseous, like helium, neon and argon. According to his proposal, a (positive) exponential term accounts for the repulsion due to the overlapping of the closed electron shells in these atoms\cite{Buckingham}. This potential has also been extensively used in simulations of molecular dynamics.

\begin{table}
\label{Fit}
\caption{Potential parameters from Eq. (\ref{Potencial}) for several $\gamma$. Negative (positive) $A$ yields attraction (repulsion) between the layers. [$\chi^2$ accounts for the mean square error for every interpolating function.]}
\centering
\begin{tabular}{| c | c | c | c | c | c |}
 \hline  
 $\gamma$ & A & B & C & D & $\chi^2$ \\ 
 \hline
 1 & $\pm$ 0.0495 & -6.2 & 0 & 0 & 0.003\\   
 1.3 & $\pm$ 0.620 & -3.438 & -9.718 & -2.938 & 0.003\\  
 1.4 & $\pm$ 0.724 &  -3.313 & -8.880 & -2.723 & 0.003\\  
 $\sqrt{3}$ & $\pm$ 1.030 & -3.026 & -6.481 & -2.164 & 0.003\\  
 1.8 & $\pm$ 1.079 & -2.988 & -6.062 & -2.074 & 0.0005\\  
 1.9 & $\pm$ 1.016 & -3.035 & -6.013 & -2.123 & 0.0005 \\
 2 & $\pm$ 1.012 & -3.038 & -5.927 & -2.122 & 0.0005 \\ 
 \hline
\end{tabular}
\end{table}

Former results have shown that ASI layers' energetics are quite sensitive to their height separation, $h$, and their respective configuration. In the following, we shall present a brief analysis of how it may be affected if the layers are twisted each other. 

By keeping the layers in opposite configurations and separated by a height offset $h=0.5$, we can rotate them by an angle $\varphi$ to analyze how the energy is affected.  What we find is that, due to the dipolar interactions between nanoislands, the energy will be completely disrupted even for a small angle, giving a strongly diminished interaction energy. Although the behavior of the potential with the rotating angle, $V(\varphi)$, is not particularly revealing, it is instructive to pay attention to the depth of their minima for distinct aspect ratio, as depicted in  Fig. 3. In some specific angles, the dipoles meet a position in which most of the interactions are maximized, giving us sharp peaks on the energy (Fig. 3a). At such configurations, the layers seem to form a superlattice with very interesting patterns, such as stripes for $\gamma = \sqrt{3} - S1$ and elipses for $\gamma = \sqrt{2}$.
\begin{figure}
\includegraphics[width=0.99\linewidth]{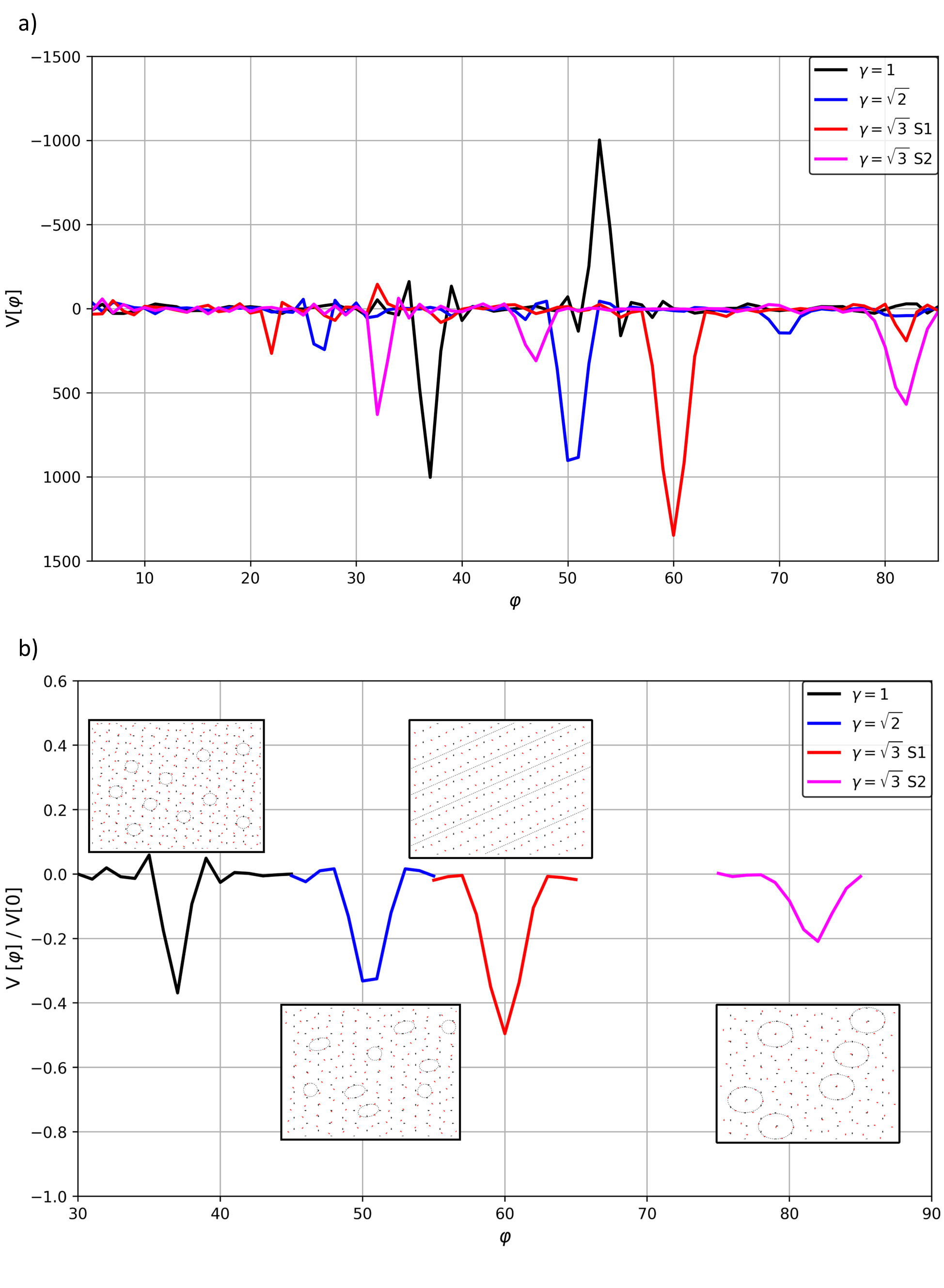}

\caption{{\bf a)} Energy behavior due to the twisting angle between the layers, $\varphi$, for a number of $\gamma$ values (each layer is composed of 841 vertices, leading to a relatively high interaction energy). {\bf b)} Normalized potential behavior around their respective minima for each $\gamma$. The deepest minima occur at $\gamma=\sqrt3$ composed by $t1$-vertex yielding stripe-like superlattice pattern (red color), while $t2$ vertices lead to elongated vortex-type moire pattern (magenta color). The square geometry, $\gamma=1$, presents a relatively deep minima with circular moire vortex pattern (black color). In general, as the vortex patterns become elongated, resembling an elliptical shape, their respective minimum become less deep.}
\end{figure}

\section{Thermodynamics}

Finally, we now investigate the basic thermodynamics features of the BRASI systems, namely considering the specific heat as a function of the temperature and of the $\gamma$-parameter, as the height offset, $h$, is varied. For that, the standard Monte Carlo technique, in which the probability of a single spin flip is ruled by the Boltzmann factor $\exp(-\Delta E/k_B T)$, is employed. Here, $\Delta E$ represents the energetic cost of spin inversion and $k_B$ is the Boltzmann constant. This criterion guaranties that the configurations of lower energy are the most likely in thermodynamic equilibrium. The specific heat is obtained by computing the energy fluctuations, $c=(\langle E^2\rangle - \langle E\rangle^2)/N k_B T^2$.
\begin{figure}[!ht]
\includegraphics[width=0.98\linewidth]{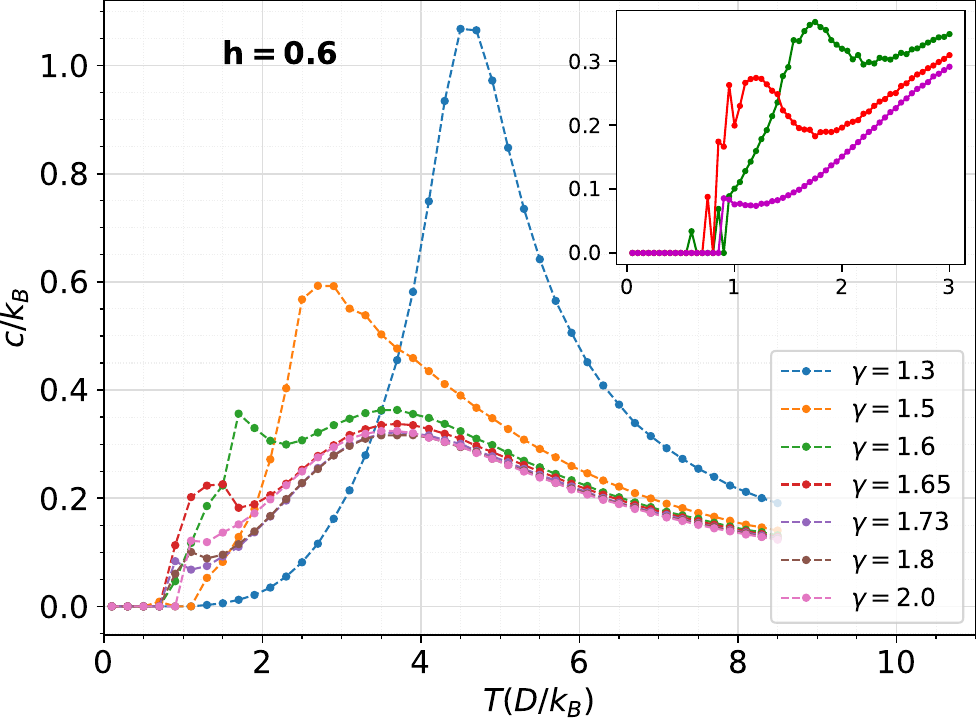} 
\caption{Specific heat as a function of temperature for different values of $\gamma$ and for fixed interlayer spacing $h/b=0.6$. BRASI systems exhibits a single specific heat peak whenever $\gamma<1.5$, whereas two peaks occur for $\gamma > 1.5$. The second peak, around $T\approx3.5~D/k_B$ for $\gamma > 1.5$, comes about due to the bilayer dynamics, namely it is traced back with a phase in which the two layers tend to be configured as the inverse of each other.
The other peaks resembles those appearing in a single layer rectangular ASI, being associated with the transition from the paramagnetic (disordered) to the spin ice phase (ordered).} 
\end{figure}

\begin{figure}[!ht]
\includegraphics[width=0.99\linewidth]{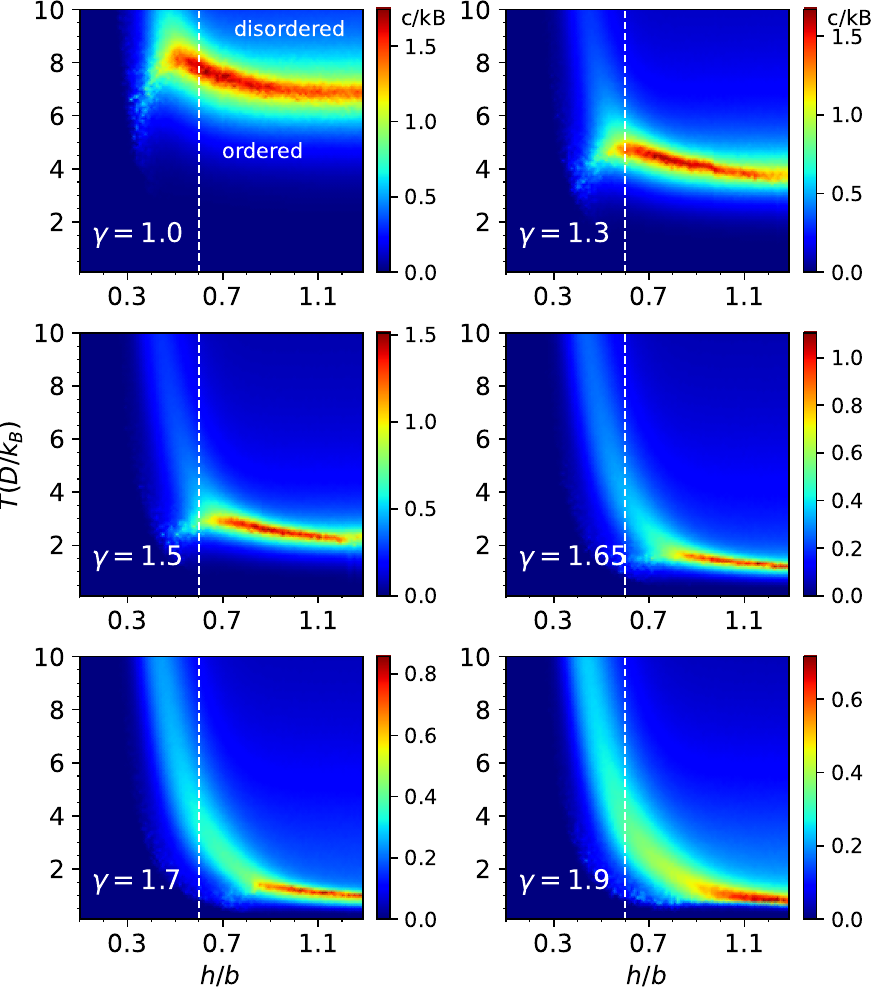} 
\caption{Contour plot of the in-plane specific heat $c/k_B$ as a function of interlayer distance $h$ and temperature. In the dark-blue region, the specific heat is essentially zero and no dynamical activity takes place. The red region indicates where the specific heat is most pronounced, corresponding to the occurrence of the order–disorder phase transition.}
\end{figure}
In general, the specific heat decreases as $\gamma$ increases. Interestingly, considering as an example $h/b=0.6$ and $\gamma>1.5$, there are two peaks in the specific heat (see Fig.4, which shows the specific heat as a function of the temperature). The weaker peak, which occurs at lower temperature (inset in Fig.4), resembles that observed in a single layer rectangular ASI \cite{Rectangular-2012} and has the usual interpretation (see Ref. \cite{Rectangular-2012, Bilayer-Fabio-21}). Indeed, this peak delimits the separation between the paramagnetic (disordered) and the spin ice (ordered) phases. However, a second peak for $\gamma > 1.5$ and $T\approx3.5~D/k_B$ emerges exclusively from the dynamics of the bilayer. It is worth noting that the specif heat peak at $T\approx3.5~D/k_B$ is almost independent of the aspect ratio $\gamma$, so it must be a characteristic of the interlayer distance. Analyzing the spin configuration at low-temperature, when dynamical activity is still present, such as charge creation, annihilation, and migration, we have observed that although the two layers have the same spin configuration, the spins of one layer are opposite to the spins of the other layer, leading to opposite effective magnetic charges in these two different parallel layers. Consequently, there is no residual charge in a BRASI system. In this intermediate regime between the two peaks of the specific heat, noticeably, two effects are competing: 1. order-disorder transition; 2. tendency for the two layers to be configured inversely to each other. Then, as the temperature increases, the system finds a new kind of phase in which the excitations surge autonomously in the layers.

To better describe BRASIs thermodynamics, we plot a diagram considering the temperature as a function of $h/b$ (see Fig. 5, where the dashed vertical white line highlights $h/b=0.6$; see also the curves in Fig. 4).
Firstly, it is worth to compare the temperature of the specific heat peak at $h/b=1$ for different values of $\gamma$. We observe that the temperature of the specific heat peak decreases monotonically as $\gamma$ increases, hitting a minimum value at $ ~ 1.5D/k_B$ for $\gamma=\sqrt{3}$. It is also worth noting that, for a given value of $\gamma$, the specific heat does not suffer substantial changes for $h/b > 1$, because the two layers are practically decoupled. This regime in which the layers are decoupled has already been previously observed in the square bilayer \cite{Bilayer-Fabio-21}.
On the other hand, for $h/b<1$, the dynamics is strongly dependent of the aspect ratio of the lattice.
Three regimes (indicated by colors) are observed: (1) the temperature of the specific heat peak increases as $h/b$ decreases (red region); (2) the specific heat suffers a strong attenuation (green region); (3) the specific heat vanishes in all intervals of temperature (dark-blue region), since the two layers are strongly coupled to enable significant spin dynamics.

\section{Conclusions and Prospects}

We have investigated two staked rectangular ASI's. Compared to the square geometry, the potential between the layers now acquires an extra exponential term, resembling a Buckingham-like interaction. Whenever one layer is twisted to the other, well-defined super-lattice patterns show up at special angles, indicating the most stable configuration of the bilayer. At a fixed height offset, $h$, the potential depth in the minima also depends on the parameter $\gamma$. At $\gamma=\sqrt3$ the deepest minimum is related to a stripe-like moire pattern. In the remaining cases, these minima coincide with vortex-type super-lattice patterns. Specific heat as a function of temperature exhibits a single peak for $\gamma < 1.5 $, whereas two peaks occur for $\gamma > 1.5 $.

Prospects for a forthcoming investigation include the role played by monopole and string-like excitation on one or both layers and how they affect the system properties as a whole. New findings may be important to highlight potential applications of ASI systems in neuromorphic prototypes for brain-inspired computation \cite{Clodoaldo,Sultana} and active inference \cite{STAMPS}.\\

\begin{acknowledgments}
The authors thank the Brazilian agencies CAPES, CNPq, FAPEMIG, INCT/CNPq - {\it Spintr\^onica e Nanoestruturas Magn\'eticas Avan\c{c}adas (INCT-SpinNanoMag)}, and {\it Rede Mineira de Nanomagnetismo/FAPEMIG}, for financial support. The authors also acknowledge the National Laboratory for Scientific Computing (LNCC/MCTI) for providing access to the Santos Dumont supercomputer, high performance computing (HPC) resources, and UFRB for facilitating this access.
\end{acknowledgments}

\section*{Data Availability Statement}

The data that supports the findings of this study
are available from the corresponding author upon
reasonable request.


\appendix*
\vskip .5cm
\centerline{--------------------------------------------------\\}
\section{}
To calculate the strength of the interaction between two $t_1$ or $t_2$ vertices, we need to determine how much residual charge or magnetic moment they bear. The coupling of $t_1$ vertices, is simply the interaction of two magnetic charges:
\begin{equation}
    E_{t1}(a,b,r_{12}) = \frac{\pm 4\mu_0 \mu^2 (b-a)^2}{4\pi r_{12} (ab)^2},
\end{equation}
whereas $t_2$ vertices couples as a pair of magnetic dipoles, like below:
\begin{equation}
\begin{split}
    E_{t2}(a,b,r_{12}) = \frac{\mu_0 \mu^2 (b+a)^2 (a^2 + b^2)}{4\pi r_{12}^3 (ab)^2}\,\times\\ \left[\pm 1-3(\hat{m}_1 \cdot \hat{r}_{12})(\hat{m}_2 \cdot \hat{r}_{12})\right].
\end{split}
\end{equation}

For simplicity, we consider that the vertices are exactly above each other, separated by a distance $r_{12} = 0.5$, so that for $\gamma = \sqrt{2}$ (i.e. $a = 1$ and $b = \sqrt{2}$) and $\gamma = \sqrt{3}$ (i.e. $a = 1$ and $b = \sqrt{3}$), one obtains:
\begin{equation}
\begin{split}
    \frac{E_{t1}(1,\sqrt{3},0.5)}{E_{t1}(1,\sqrt{2},0.5)} \approx 2.08 \\ \frac{E_{t2}(1,\sqrt{3},0.5)}{E_{t2}(1,\sqrt{2},0.5)} \approx 1.14.
\end{split}
\end{equation}
Thus, in a state composed by $t_1$ the energy doubles as one goes from $\gamma = \sqrt{2}$ to $\gamma = \sqrt{3}$, whereas it increases by only 14$\%$ if $t_2$ vertices are in order.\\
\centerline{---------------------------------------------------\\}




\end{document}